\documentclass[aps,prd,12pt,nofootinbib,preprintnumbers,eqsecnum,showpacs]{revtex4}

\usepackage{amssymb,amsmath,amsfonts,amsbsy,graphicx}

\def\mathbi#1{\textbf{\em #1}}

\newcommand{\mpl}{m_{\rm Pl}}
\newcommand{\fnl}{f_{\rm NL}}
\newcommand{\gnl}{g_{\rm NL}}
\newcommand{\tnl}{\tau_{\rm NL}}
\newcommand{\eff}{{\rm eff}}
\newcommand{\calH}{{\cal H}}

\newcommand{\calO}{{\cal O}}
\newcommand{\calP}{{\cal P}}
\newcommand{\calR}{{\cal R}}

\def\M{M_\mathrm{eff}}
\def\m{m_\mathrm{eff}}
\def\bp{\mathbi{p}}
\def\bpsum{\mathbi{p}_1+\mathbi{p}_2+\mathbi{p}_3}
\def\psum{p_1+p_2+p_3}

\def\pprod{p_1p_2p_3}

\begin{document}

\preprint{APCTP-Pre2013-007, YITP-13-46}

\title{Equilateral non-Gaussianity from heavy fields}

\author{
Jinn-Ouk Gong,$^{1,2}$\footnote{jinn-ouk.gong\_AT\_apctp.org}
Shi Pi$^{1}$\footnote{spi\_AT\_apctp.org}
and
Misao Sasaki$^3$\footnote{misao\_AT\_yukawa.kyoto-u.ac.jp}
}

\affiliation{
~\\
$^1$Asia Pacific Center for Theoretical Physics, Pohang 790-784, Korea
\vspace{3mm}
\\
$^2$Department of Physics, Postech, Pohang 790-784, Korea
\vspace{3mm}
\\
$^3$Yukawa Institute for Theoretical Physics, Kyoto University, Kyoto 606-8502, Japan
}

\date{\today}

\begin{abstract}
The effect of self-interactions of heavy scalar fields during inflation on the
primordial non-Gaussianity is studied. We take a specific constant-turn
quasi-single field inflation as an example. We derive an effective theory with
emphasis on non-linear self-interactions of heavy fields and calculate
the corresponding non-Gaussianity, which is of equilateral type and can be as
relevant as those computed previously in the literature.
We also derive the non-Gaussianity by directly using the in-in formalism,
and verify the equivalence of these two approaches.

\end{abstract}


\maketitle

\newpage

\section{Introduction}
\label{sec:intro}

Inflation is the most successful paradigm among the models of the early universe~\cite{inflation}. According to this picture, the universe has suffered a period of accelerated expansion which can solve the traditional puzzles as the horizon problem. Also, during inflation the quantum fluctuations generated deep inside the horizon are stretched to the super-horizon scales and then frozen, which later become the seed for the large scale structure of the universe we can observe today~\cite{Weinberg:2008zzc}. The observations by the most recent PLANCK satellite show that the primordial curvature perturbation has nearly scale invariant spectrum and follows almost perfect Gaussian statistics~\cite{planck2013}. Therefore, any tiny deviation from these observations, if ever detected, tells us important properties of the curvature perturbation and the physics behind it~\cite{Ade:2013nlj}.

Among them, non-Gaussianity is a promising probe that can be used to distinguish the zoology of inflationary models~\cite{nGreviews}. Although the recent PLANCK data is consistent with vanishing non-Gaussianity at 1$\sigma$ confidence level~\cite{Ade:2013ydc}, still it may be detected at smaller scales, e.g. in the large scale structure~\cite{Desjacques:2010jw}. Also, given the prior different from featureless power spectrum, it is questionable whether the estimates on non-Gaussianity, in particular of equilateral shape, are still consistent with zero. Thus it would be too premature to close any further study on non-Gaussianity.

It is known that general single field inflation with non-trivial speed of sound $c_s$ generates such non-Gaussianity~\cite{Chen:2006nt}. The action of general single field inflation may be the low energy realization of a parent theory, whose additional mass scales are manifest in the low energy effective theory through the couplings that parametrize the derivative expansion~\cite{Achucarro:2012sm}. The effective theory approach is particularly efficient when the mass hierarchies are large~\cite{masshierarchy}. That is, except for the almost massless inflation, there are many scalar fields whose masses are much larger than that of the inflaton. In that case, we can systematically integrate out the heavy fields, and the resulting effective action is that of a single field with non-trivial $c_s$~\cite{Tolley:2009fg}. Explicitly, if we have started from a canonical two-field system\footnote{See~\cite{Cespedes:2013rda} for a model with two heavy fields.}, the power spectrum is given by that of k-inflation~\cite{k-inflation},
\begin{equation}\label{PowerSpec}
\calP_\calR = \frac{H^2}{8\pi^2\mpl^2\epsilon c_s} \, ,
\end{equation}
where $c_s^{-2} = 1+4\dot\theta_0^2/\M^2$ with $\dot\theta_0$ being the angular velocity of the inflaton trajectory in the field space and $\M$ being the effective mass of the heavy field~\cite{Achucarro:2012sm,Achucarro:2010da,Pi:2012gf}. We can see that this effective sound speed reflects the correction to the power spectrum due to the heavy field intermediation. Likewise, non-Gaussianity can be written in terms of $c_s$ as $\fnl \sim 1/c_s^2$~\cite{Chen:2006nt}.

Note, however, that here non-linear interactions of the heavy fields are not included.
This point can be stated more clearly from the action for the Goldstone boson $\pi$ of gravity in a de Sitter background~\cite{Cheung:2007st}:
\begin{equation}
\label{ActionGoldstone}
S_\pi = \int~d^4x a^3 \left\{ -\mpl^2\dot{H} \left[ \dot\pi^2 - \frac{(\nabla\pi)^2}{a^2} \right] + 2M_2^4\dot\pi \left[ \dot\pi + \dot\pi^2 - \frac{(\nabla\pi)^2}{a^2} \right] - \frac{4}{3}M_3^4\dot\pi^3 + \cdots \right\} \, ,
\end{equation}
where, when we solve the equation of motion of the heavy field linearly
[see (\ref{eom:DeltaSigma}) and below], ignoring possible nonlinear interactions,
the sound speed $c_s$ and the couplings $M_n^4$ are uniquely related
by~\cite{Achucarro:2012sm}
\begin{equation}\label{Mn}
M_n^4 = (-1)^nn!|\dot{H}|\mpl^2 \left( \frac{c_s^{-2}-1}{4} \right)^{n-1} \, ,
\end{equation}
so that $M_3^4 \sim \M^{-4}$, and especially the pure cubic interaction terms
of the heavy field would result in a contribution suppressed by $\M^{-6}$,
thus at first look it seems their contributions to non-Gaussianity are negligible.

However, it is possible that non-linear self-interactions of the heavy field may
be important. A simple example is the quasi-single field inflation~\cite{qsfi},
where the non-linear interaction of the heavy field is not suppressed by the
 slow-roll parameters.
In this case, the non-linear self-interaction of the isocurvature perturbation will
generate the main contribution to non-Gaussianity when the isocurvature mass $\M$
is of $\calO(H)$.

A mass parameter $\nu=\sqrt{9/4-\M^2/H^2}$ is used to describe the size and
shape of non-Gaussianity: when $\M$ is very small ($\nu\to3/2$) the shape of
non-Gaussianity is purely local, meanwhile when $\M$ approaches $3H/2$ ($\nu\to0$)
 the shape changes to ``intermediate'' which lies between local and equilateral.
 That is, as $\M$ increases, the non-Gaussianity generated by its self-interactions
becomes more and more equilateral.
This equilateralization process continues as $\M$ exceeds $3H/2$, which corresponds
to a pure imaginary mass parameter $\nu=i\mu=i\sqrt{\M^2/H^2-9/4}$.
Finally, as $\M$ becomes infinitely large ($\nu\to i\infty$), it is expected
that the shape will become purely equilateral. This will be explicitly
shown in this article.

In the limit $\M\to\infty$ the ``quasi-single field'' inflation becomes
``single field'' inflation as the heavy isocurvature mode is completely
frozen by the time when the wavelength of the fluctuations exceeds the Hubble
horizon size. Thus it seems clear that the quasi-single field inflation under this
condition can also be approximated by an effective single field theory.
In~\cite{Pi:2012gf}, it was shown that $\calP_\calR$ in this model when $\M$ is
large is exactly the same as that of the corresponding effective single
field (see also~\cite{Chen:2012ge}). Therefore, it is also interesting to
see whether the effective single field approach and a direct calculation
using the in-in formalism~\cite{In-in} gives the same result for non-Gaussianity.

To include self-interactions of the heavy field, we have to extend
the effective single field method to non-linear level. In this article, we will
calculate the bispectrum from the self-interactions of the heavy field by using
 both approaches. The result is that the non-Gaussianity originated from the
self-interaction of the heavy field is purely equilateral, as expected, and
both the size and shape obtained by the two methods are the same.
Besides, it can dominate non-Gaussianity, and is even possible to be large.

This article is organized as followed. In Section~\ref{sec:EFT} we use the effective field theory to study the quasi-single field inflation with large mass, focusing on the non-Gaussianity generated by the term corresponding to the heavy-field interactions. In Section~\ref{sec:in-in} we turn to the in-in formalism to calculate non-Gaussianity and compare the results. We discuss shortly the possible magnificence of the underline physical effects of the hidden heavy sector of inflation and then conclude in Section~\ref{sec:discuss}.

\section{Effective field theory approach}
\label{sec:EFT}

It is known that an inflationary model of multiple scalar fields can be described by a single field, if the masses of the scalar fields except for the lightest one, usually identified as the inflaton, are all very large. Such an approach provides us a prospect to connect the single-field and multi-field inflation models in the weakly coupled regime~\cite{weaklycoupled}. The main result of the previous efforts is that, when the field trajectory takes a turn in the field space, the curvature perturbation is equivalent to that in general single field inflation with an effective speed of sound $c_s$, which is connected to the angular speed of the trajectory and the heavy masses. In this section we will study the bispectrum by adopting the effective theory approach, emphasizing the effects from the
self-interaction of heavy fields.

We take as an example a well-developed toy model, the quasi-single field inflation~\cite{qsfi}. It can mimic the process of a segment of the inflationary trajectory when it undergoes a slow-roll turn in the field space. To be specific, we consider a motion along an arc with the radius $R$ in the field space. And the fields are decomposed into the light field $R\theta$ and the massive one $\sigma$, along and perpendicular to the trajectory respectively~\cite{Burgess:2012dz}. The matter Lagrangian for the quasi-single field inflation in a constant turn case is
\begin{equation}
\label{ActionQSI}
S_m = \int d^4x \sqrt{-g} \left[ -\frac{1}{2}\left( \tilde R+\sigma \right)^2(\partial_\mu\theta)^2 - \frac{1}{2} (\partial_\mu\sigma)^2 - V_\text{sr}(\theta) - V(\sigma) \right] \, ,
\end{equation}
where $\tilde R$ is the radius of the circular bottom of the trough in the field space, and $\left(\tilde R+\sigma\right)\theta$ and $\sigma$ are the tangential and radial fields along and perpendicular to the classical field trajectory, respectively. The potential is assumed to be ``seperable'', with $V_\mathrm{sr}(\theta)$ representing the slow-roll potential along the potential trough, and $V(\sigma)$ being the potential of $\sigma$ only. The equations of motion for the classical trajectory are
\begin{align}
3\mpl^2H^2 & = \frac{1}{2}R^2\dot\theta_0^2+V_\mathrm{sr}+V \, ,
\\
-2\mpl^2\dot H & = R^2\dot\theta_0^2 \, ,
\\
0 & = R^2\ddot\theta_0+3R^2H\dot\theta_0+V_\mathrm{sr}' \, ,
\\
0 & = V'-R\dot\theta_0^2 \, .
\end{align}
In the above we have defined $R\equiv\tilde R+\sigma_0$ with $\sigma_0$ being a constant radial displacement from the bottom of the trough. We can define the slow-roll parameters as
\begin{align}
\label{def:ep}
\epsilon &\equiv - \frac{\dot H}{H^2} = \frac{R^2\dot\theta_0^2}{2\mpl^2H^2} \approx \frac{\mpl^2}{2} \left( \frac{V_\mathrm{sr}'}{R V_\mathrm{sr}}\right)^2 \, ,
\\
\label{def:eta}
\eta &\equiv \frac{\dot\epsilon}{H\epsilon} \approx -2\mpl^2 \frac{V_\mathrm{sr}''}{R^2 V_\mathrm{sr}} + 2\mpl^2\left( \frac{V_\mathrm{sr}'}{R V_\mathrm{sr}}\right)^2 \, .
\end{align}

\begin{figure}
  \centering
  \includegraphics[width=0.4\textwidth]{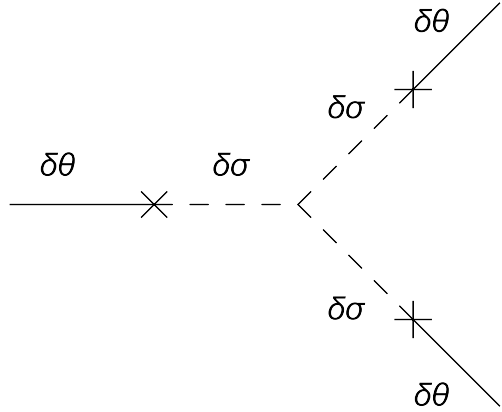}
  \caption{A schematic diagram that shows how the self-interaction of the heavy field $\delta\sigma$ affects the three-point function of the light field $\delta\theta$. The contribution of the cubic interaction $V'''\delta\sigma^3/6$ is transferred to three $\delta\theta$'s by the quadratic interaction $2R\dot\theta_0\dot{\delta\theta}\delta\sigma$.}
  \label{fig:3vertex}
\end{figure}

Let us investigate the perturbative property of (\ref{ActionQSI}). We adopt the spatially flat gauge and define
\begin{align}
\theta(t,\mathbi{x}) & = \theta_0(t)+\delta\theta(t,\mathbi{x}) \, ,
\\
\sigma(t,\mathbi{x}) & = \sigma_0+\delta\sigma(t,\mathbi{x}) \, .
\end{align}
And the potential $V(\sigma)$ can be expanded around the classical solution $\sigma_0$ by
\begin{equation}\label{Vseries}
V(\sigma)=V(\sigma_0)+V'(\sigma_0)\delta\sigma+\frac{1}{2}V''(\sigma_0)\delta\sigma^2+\frac{1}{6}V'''(\sigma_0)\delta\sigma^3+\cdots \, .
\end{equation}
Then the perturbed action, with the gravity sector being neglected, is up to the third order given by
\begin{align}\label{ActionPert}
S[\delta\theta,\delta\sigma] & = \int dt d^3x a^3 \left[ \frac{1}{2} R^2 \dot {\delta\theta}^2 - \frac{R^2}{2a^2}(\nabla\delta\theta)^2 + \frac{1}{2}\dot{\delta\sigma}^2 - \frac{1}{2a^2}(\nabla\delta\sigma)^2 - \frac{1}{2}\m^2\delta\sigma^2+2R\dot\theta_0\dot{\delta\theta}\delta\sigma\right.
\nonumber\\
& \hspace{2.5cm} \left. + R\delta\sigma\dot{\delta\theta}^2 +\dot\theta_0\dot{\delta\theta}\delta\sigma^2 - \frac{R}{a^2}\delta\sigma\left(\nabla\delta\theta\right)^2 - \frac{1}{6}V'''(\sigma_0)\delta\sigma^3 +\cdots\right] \, ,
\end{align}
where $\m^2 = V''(\sigma_0)-\dot\theta_0^2$. In the action above we have only preserved the leading order terms in slow-roll parameters. Our interest is the last term, which contributes to the three-point function of $\delta\theta$ as shown in Figure~\ref{fig:3vertex}.

Now we adopt the effective single field theory approach to compute the bispectrum. The equation of motion for $\delta\sigma$ is found from (\ref{ActionPert}) as, with $\nabla^2\equiv\delta^{ij}\partial_i\partial_j$,
\begin{equation}\label{eom:DeltaSigma}
\ddot{\delta\sigma} + 3H\dot{\delta\sigma} - \left( \frac{\nabla^2}{a^2} - \m^2 \right)\delta\sigma+\frac{V'''}{2}\delta\sigma^2 = 2R\dot\theta_0\dot{\delta\theta}\,.
\end{equation}
In the case when the effective mass of $\delta\sigma$ is very large, the term $\m^2\delta\sigma$ will dominate the equation. Therefore, we can neglect the space-time derivatives of $\delta\sigma$ in (\ref{eom:DeltaSigma}) and find the solution of (\ref{eom:DeltaSigma}) perturbatively
\footnote{The solution to (\ref{Constraint}) up to arbitrary order $n$ and the contribution to $n$-spectra are briefly discussed in Appendix \ref{app:n-point}.}
\begin{equation}\label{Constraint}
\delta\sigma = \frac{2R\dot\theta_0}{\m^2}\dot{\delta\theta}+\left(\frac{R}{\m^2c_s^2}-\frac{2R^2\dot\theta_0^2}{\m^2}\frac{V'''}{\m^4}\right)\dot{\delta\theta}^2+\cdots \, ,
\end{equation}
which is essentially a constraint, viz. $\delta\sigma$ is no more a degree of freedom but a Lagrangian multiplier. Note that to make the term $\m^2\delta\sigma$ dominant in the left hand side of (\ref{eom:DeltaSigma}), we should impose
\begin{equation}
\frac{V'''}{2}\delta\sigma\lesssim\m^2 \, ,
\end{equation}
which gives us the only constraint on the magnitude of $V'''$. We will resume the discussion on this issue at the end of this section.

Substituting (\ref{Constraint}) back into (\ref{ActionPert}), we obtain the effective single field action for $\delta\theta$ as
\begin{align}\label{ActionEff1}
S_\text{eff}[\delta\theta] & = \int dt d^3x a^3 \left[ \frac{1}{2} R^2\dot{\delta\theta}^2 \left( 1+4\frac{\dot\theta_0^2}{\m^2} \right) - \frac{R^2}{2a^2}\left(\nabla\delta\theta\right)^2 \right.
\nonumber\\
& \hspace{2.5cm} \left. + \left( \frac{2R^2\dot\theta_0^2}{\m^2}+\frac{4R^2\dot\theta_0^3}{\m^4}-\frac{4R^3\dot\theta_0^3}{3\m^6}V''' \right) \dot{\delta\theta}^3 - \frac{2R^2\dot\theta_0}{a^2\m^2}\dot{\delta\theta}\left(\nabla\delta\theta\right)^2 \right] \, .
\end{align}
We see from this action that, if we define an effective sound speed as
\begin{equation}\label{def:cs-Re}
\frac{1}{c_s^2} \equiv 1 + \frac{4\dot\theta^2}{\m^2} \, ,
\end{equation}
(\ref{ActionEff1}) is equivalent to that of general single field inflation~\cite{Chen:2006nt,k-inflation}. To evaluate the observable quantities, we have to transfer the action into that of the curvature perturbation. It is known that the curvature perturbation on the comoving slices $\calR$ is given in terms of the field fluctuation on the flat slices along the trajectory $\delta\theta$ as~\cite{smvariable}
\begin{equation}\label{def:CurvPert}
\calR = -\frac{H}{\dot\theta_0}\delta\theta \, .
\end{equation}
Doing some integrations by parts and neglecting some slow-roll suppressed terms, we find
\begin{align}\label{ActionEff2}
S_\mathrm{eff}[\calR] & = \mpl^2\int dt d^3x~a^3 \left\{ \frac{\epsilon}{c_s^2} \left[ \dot{\cal R}^2 - c_s^2 \frac{(\nabla\mathcal R)^2}{a^2} \right] \right.
\nonumber\\
& \hspace{3.5cm} - \frac{H^2\epsilon}{c_s^2}\left[ \left(\frac{1}{c_s^2}-1\right) - \frac{c_s^2}{2}\left(\frac{1}{c_s^2}-1\right)^2 - c_s^2\frac{RV'''}{6\m^2}\left(\frac{1}{c_s^2}-1\right)^2 \right] \frac{\dot{\mathcal R}^3}{H^3}
\nonumber\\
& \hspace{3.5cm} \left. + \epsilon \left(\frac{1}{c_s^2}-1\right) \frac{\dot\calR}{H}\frac{\left(\nabla\mathcal R\right)^2}{a^2} + \cdots \right\} \, .
\end{align}
We can now compare the effective action of the Goldstone boson (\ref{ActionGoldstone}) with (\ref{ActionEff2}) to determine the undetermined coefficients $M_2^4$ and $M_3^4$, and find
\begin{align}
M_2^4 & = \frac{1}{2}\epsilon\mpl^2 H^2\left(\frac{1}{c_s^2}-1\right) \, ,
\\\label{M4}
M_3^4 & = \frac{3}{4}\epsilon\mpl^2H^2\left(\frac{1}{c_s^2}-1\right)^2\left(\frac{RV'''}{6\m^2}-\frac{1}{2}\right) \, .
\end{align}
Comparing (\ref{M4}) with the previous relation (\ref{Mn}) with $n=4$, we see the origin of the parameter $M_3$ of the effective field theory of inflation that was not found before: it is the reflection of the non-linear self-interaction(s) of the heavy field(s) during inflation. This statement based on the fact that the $V'''$term will dominate $M_3$, which, as well as the constraint of its magnitude mentioned before, will be shown later at the end of this section.

Non-Gaussianity associated with the perturbative action (\ref{ActionEff2}) is well known: in the regime of our interest $\nu\to i\infty$ the bispectrum is of equilateral shape, and especially the contribution from $\dot{\mathcal R}^3$ gives, with $S_\lambda$ being the corresponding shape function,
\begin{align}
\label{def:Shape}
\langle\mathcal{R}(\mathbi{p}_1)\mathcal{R}(\mathbi{p}_2)\mathcal{R}(\mathbi{p}_3)\rangle = & (2\pi)^7 \delta^{(3)}(\bpsum)\mathcal{P}_\mathcal{R}^2\frac{S_\lambda}{(\pprod)^2} \, ,
\\
\label{ShapeEFTbyMass}
S_\lambda = & -\frac{6\dot\theta_0^2}{\m^2}\left[1-\frac{2\dot\theta_0^2}{\m^2}c_s^2\left(1+\frac{RV'''}{3\m^2}\right)\right]\frac{p_1p_2p_3}{\left(p_1+p_2+p_3\right)^3} \, .
\end{align}
The non-linear parameter $\fnl=(10/9)S(p_1=p_2=p_3)$~\cite{Chen:2010xka} is given by
\begin{align}
\fnl^\lambda & = -\frac{20}{81}\frac{\dot\theta_0^2}{\m^2}\left[1-\frac{2\dot\theta_0^2}{\m^2}c_s^2\left(1+\frac{RV'''}{3\m^2}\right)\right]
\nonumber\\
&=-\frac{20}{81}\left(\frac{\dot\theta_0}{\m}\right)^2+\frac{40}{81}c_s^2\left(\frac{\dot\theta_0}{\m}\right)^4 + \frac{40}{243}\frac{RV'''}{\m^2}c_s^2\left(\frac{\dot\theta_0}{\m}\right)^4 \, .
\label{fNLEFTbyMassSeries}
\end{align}
It will be more clear if we use $\m$ to represent $c_s$, but we will preserve it as it originates from the definition of power spectrum in (\ref{def:Shape}). This is to distinguish the $c_s$ from the definition of effective mass, which will be useful in Section \ref{sec:in-in}. Note that the effective single field description is working when the mass of the isocurvature perturbation is very large, i.e.
\begin{equation}
  \left(\frac{H}{\m}\right)^2 \ll 1 \, .
\end{equation}
There is no constraint on either $\dot\theta_0/H$ or especially $\dot\theta_0/\m$~\cite{masshierarchy}, provided that the adiabaticity condition $\big|\ddot\theta_0/\dot\theta_0\big| \ll \m$ is satisfied~\cite{Achucarro:2012sm,Cespedes:2012hu} although the quasi-single field inflation with a large $\dot\theta_0/H$ will be quite unnatural. In Section~\ref{sec:in-in}, $\dot\theta_0/H$ will be set to be small to make the in-in formalism valid, so in this section we also implicitly assume $\dot\theta_0/H\ll1$.

As the main contribution to the equilateral non-Gaussianity is displayed in (\ref{fNLEFTbyMassSeries}), an important question is which one will dominate. The non-Gaussianity originated from the self-interaction of the heavy field will overwhelm the others which are suppressed by slow-roll parameters. This is clearly true when $\m\lesssim H$, as claimed in~\cite{qsfi}. But now as $\m$ is very large, the term proportional to $V'''$ is suppressed by an additional $\m^{2}$ factor. To convince ourselves this term is still or even more important, we should estimate how large it could be. We begin by imposing the condition under which the perturbative series of the potential (\ref{Vseries}) converges,
\begin{equation}\label{est:pertvalid}
\frac{V'''\delta\sigma}{3V''}=\frac{V'''\delta\sigma}{3\left(\m^2+\dot\theta_0^2\right)}\lesssim1 \, .
\end{equation}
Now we use the constraint (\ref{Constraint}) with (\ref{def:ep}) to obtain
\begin{equation}\label{est:DeltaSigma}
\sigma = 2\sqrt{2\epsilon}\frac{H\mpl}{\m^2}\dot{\delta\theta} \, .
\end{equation}
To estimate $\dot{\delta\theta}$, from the conservation of $\calR$ on large scales we can derive
\begin{equation}\label{est:DeltaTheta}
\dot{\delta\theta} = \frac{H\eta}{2}\delta\theta \sim \frac{H^2\eta}{4\pi Rc_s^{1/2}} \, .
\end{equation}
Substituting (\ref{est:DeltaTheta}) into (\ref{est:DeltaSigma}), we have
\begin{equation}
\delta\sigma \sim \sqrt{\frac{2\epsilon}{c_s}}\eta\frac{H^3\mpl}{2\pi \m^2R} \, .
\end{equation}
Thus we can estimate $V'''$ as
\begin{equation}
V'''  \sqrt{\frac{c_s}{2\epsilon}}\frac{6\pi \m^2 \left( \m^2+\dot\theta_0^2 \right)R}{\eta H^3\mpl} \, .
\end{equation}
This relation validates the power series of (\ref{Vseries}) as well as the perturbative solution of (\ref{Constraint}). Therefore we can estimate the contribution to $\fnl$ due to the self-interactions of the heavy field, i.e. the $V'''$ term in (\ref{fNLEFTbyMassSeries}), by
\begin{equation}
\frac{RV'''\dot\theta_0^4}{3\m^4\left(\m^2+4\dot\theta_0^2\right)} \lesssim
\frac{4\pi}{\eta}\sqrt{\frac{c_s\epsilon}{2}}\frac{\mpl}{H}\left(\frac{\dot\theta_0}{\m}\right)^2\frac{\m^2+\dot\theta_0^2}{\m^2+4\dot\theta_0^2} \sim \frac{1}{\eta\mathcal{P}_\mathcal{R}^{1/2}}\left(\frac{\dot\theta_0}{\m}\right)^2 \, .
\end{equation}
Comparing to the other terms in (\ref{fNLEFTbyMassSeries}), we see that indeed the $V'''$ term will dominate $\fnl$ because of the $\eta^{-1}\mathcal{P}_\mathcal{R}^{-1/2}$ factor. Although, as we have mentioned above, a natural model requires $\dot\theta_0^2/\m^2$ to be small, we can still have a large $\fnl$ because of this prefactor.

\section{In-in formalism calculations}
\label{sec:in-in}

In the previous section, we have adopted the effective single field theory approach to compute the contribution to non-Gaussianity of the self-interaction of heavy fields, with the quasi-single field inflation as an explicit example. In this section we use the in-in formalism~\cite{In-in}, and verify the result (\ref{fNLEFTbyMassSeries}) in the last section. We start with the action (\ref{ActionPert}). The first task is to define the canonical conjugate momenta $\pi_{\delta\theta}=\partial\mathcal{L}/\partial(\dot{\delta\theta})$ and $\pi_{\delta\sigma}=\partial\mathcal{L}/\partial(\dot{\delta\sigma})$. Then we define the Hamiltonian density by $\mathcal{H}=\pi_{\delta\theta}\dot{\delta\theta}+\pi_{\delta\sigma}\dot{\delta\sigma}-\mathcal L$ where $\dot{\delta\theta}$ and $\dot{\delta\sigma}$ are expressed by $\pi_{\delta\theta}$, $\pi_{\delta\sigma}$, $\delta\theta$ and $\delta\sigma$. We divide $\mathcal H$ into the free-field part $\mathcal{H}_0$ and two interaction parts $\mathcal{H}^I_2$ and $\mathcal{H}^I_3$, and then replacing $\pi_{\delta\theta}$ and $\pi_{\delta\sigma}$ by $\pi^I_{\delta\theta}$ and $\pi^I_{\delta\sigma}$ which satisfy the Hamiltonian equation of the free-field Hamiltonian:
\begin{align}
\label{def:ConjTheta}
\dot{\delta\theta}_I & = \left.\frac{\partial\mathcal H_0}{\partial\pi_{\delta\theta}}\right|_{\pi_{\delta\theta}=\pi_{\delta\theta}^I} \, ,
\\
\label{def:ConjSigma}
\dot{\delta\sigma}_I & = \left.\frac{\partial\mathcal H_0}{\partial\pi_{\delta\sigma}}\right|_{\pi_{\delta\sigma}=\pi_{\delta\sigma}^I} \, .
\end{align}
The last step is to use (\ref{def:ConjTheta}) and (\ref{def:ConjSigma}) to cancel $\pi_{\delta\theta}^I$ and $\pi_{\delta\sigma}^I$. We then obtain the Hamiltonian expressed in the interaction picture:
\begin{align}
\label{H0}
\calH_0 &= a^3 \left[ \frac{1}{2} R^2 \dot {\delta\theta_I}^2 +
  \frac{R^2}{2a^2}
  (\nabla \delta\theta_I)^2
+ \frac{1}{2} \dot{\delta\sigma_I}^2 + \frac{1}{2a^2} (\nabla
\delta\sigma_I)^2 + \frac{1}{2} \M^2 \delta\sigma_I^2
\right] \, ,
\\
\label{CH2}
\calH^I_2 & = - 2 R \dot\theta_0a^3 \delta\sigma_I\dot{\delta\theta_I} \, ,
\\
\label{CH3}
\calH^I_3 &=  -a^3R\delta\sigma_I\dot{\delta\theta_I}^2
-a^3\dot\theta_0\dot{\delta\theta_I}\delta\sigma_I^2
+aR\delta\sigma_I\left(\nabla\delta\theta_I\right)^2
+\frac{a^3}{6}V'''\delta\sigma_I^3 \, ,
\\
\M^2&= V'' + 3\dot \theta_0^2 \, .
\end{align}
Here we have to notice that the definition of the effective mass $\M=V''+3\dot\theta_0^2$ for the heavy isocurvature perturbation in the interaction picture $\delta\sigma_I$ is {\em different} from $\m=V''-\dot\theta_0^2$ in the Heisenberg picture: they are related by $c_s^2=\m^2/\M^2$. However, in quasi-single field inflation the difference between the masses is not essential since $c_s^{-2}-1\ll1$.

In the interaction picture, we quantize the Fourier components $\delta\theta_I(\mathbi{k})$ and $\delta\sigma_I(\mathbi{k})$ of the free fields $\delta\theta_I$ and $\delta\sigma_I$,
\begin{align}
\delta\theta_I(\bp) &= u_p a_\bp + u_p^* a_{-\bp}^\dagger \, ,
\\
\delta\sigma_I(\bp) &= v_p b_\bp + v_p^* b_{-\bp}^\dagger \, ,
\end{align}
where $a_\bp$ and $b_\bp$ are the annihilation operators of $\delta\sigma_I(\bp)$ and $\delta\theta_I(\bp)$ respectively, and each satisfies the canonical commutation relation,
\begin{align}
\left[a_\bp,a_{-\bp'}^\dagger\right] & = (2\pi)^3 \delta^3 (\bp+\bp') \, ,
\\
\left[b_\bp,b_{-\bp'}^\dagger\right] & = (2\pi)^3 \delta^3 (\bp+\bp') \, ,
\end{align}
with all the other commutators vanishing. The mode functions $u_p$ and $v_p$ satisfy the linear equations of motion followed from the free-field Hamiltonian $\calH_0$,
\begin{align}
\frac{d^2u_p}{d\tau^2} - \frac{2}{\tau} \frac{du_p}{d\tau} + p^2 u_p &=0 \, ,
\label{modefun_u}
\\
\frac{d^2v_p}{d\tau^2} - \frac{2}{\tau} \frac{dv_p}{d\tau} + \left( p^2 + \frac{\M^2}{H^2 \tau^2} \right) v_p &= 0 \,,
\label{modefun_v}
\end{align}
where $\tau=\int dt/a \approx -1/(aH)$ is the conformal time. The solutions to (\ref{modefun_u}) and (\ref{modefun_v}) are given by linear combinations of the Hankel functions of first and second kind. Requiring that the solutions approach those in the Minkowski positive frequency functions, we obtain
\begin{align}
u_p & = \frac{H}{R\sqrt{2p^3}} ( 1+i p\tau)e^{-i p\tau} \, ,
\label{mode_u}
\\
v_p & = -i e^{ -\frac{\pi}{2}\mu + i\frac{\pi}{4}} \frac{\sqrt{\pi}}{2} H  (-\tau)^{3/2} H^{(1)}_{i\mu} (-p\tau) \, ,
\label{mode_v2}
\end{align}
where $\mu^2 = \M^2/H^2-9/4 > 0$. In this article we will only concentrate on this large mass case. The contribution of the cubic interaction Hamiltonian of our interest,
\begin{equation}
H_3^I = \int d^3x\calH_3^I = \frac{V'''a^3}{6}\int\frac{d^3q_1d^3q_2d^3q_3}{(2\pi)^6} \delta^{(3)}(\mathbi{q}_1+\mathbi{q}_2+\mathbi{q}_3) \delta\sigma_I(\mathbi{q}_1)\delta\sigma_I(\mathbi{q}_2)\delta\sigma_I(\mathbi{q}_3) \, ,
\end{equation}
is transferred into three light fields in external legs by the two-point interaction Hamiltonian
\begin{equation}
H_2^I=\int d^3x\calH^I_2=-2R\dot\theta a^3\int\frac{d^3q_1d^3q_2}{(2\pi)^3} \delta^{(3)}(\mathbi{q}_1+\mathbi{q}_2) \delta\theta_I(\mathbi{q}_1)\delta\sigma_I(\mathbi{q}_2) \, .
\end{equation}
So the three-point function of the curvature perturbation is, with $H_I = H_2^I+H_3^I$,
\begin{align}
\left\langle\mathcal R^3\right\rangle & =-\left(\frac{H}{\dot\theta_0}\right)^3\left\langle\delta\theta^3\right\rangle \, ,
\\\label{def:DeltaTheta^3}
\left\langle\delta\theta^3\right\rangle & = \left\langle0\left|\left[\bar T\exp\left(i\int^t_{t_0}dt'H_I(t')\right)\right]
\delta\theta_I^3\left[T\exp\left(-i\int^t_{t_0}dt'H_I(t')\right)\right]\right|0\right\rangle \, .
\end{align}
Expanding the exponent up to the second order\footnote{The linear order terms give slow-roll suppressed contributions, i.e. the terms in (\ref{fNLEFTbyMassSeries}) except for the $V'''$-term.} and doing the contractions, we can have the three-point function of $\delta\theta$ as the sum of ten terms, $\langle\delta\theta^3\rangle = \sum_{i=1}^{10} \langle\delta\theta^3\rangle_i$. The complete form of the ten integrals are given in Appendix~\ref{app:Integral}.

These terms can be categories into two groups. One half of them, (\ref{Integral01}), (\ref{Integral02}), (\ref{Integral04}), (\ref{Integral05}) and (\ref{Integral06}), are proportional to an integral of a product of two mode functions on the entire history of inflation,
\begin{align}\label{IntegralHankelExpExact}
&\int^0_{-\infty}d\tau(-\tau)^{-1/2}H_{i\mu}^{(1)}(-p\tau)e^{-ip\tau}
\nonumber\\
&=2(-\tau)^{1/2}\left\{\left(-\frac{2}{p\tau}\right)^{i\mu}\frac{\cot\pi\mu-1}{(2i\mu-1)\Gamma(1-i\mu)}~_2F_2\left(
                                                                                                    \begin{array}{cc}
                                                                                                      1/2-i\mu, & 1/2-i\mu; \\
                                                                                                      3/2-i\mu, & 1-2i\mu; \\
                                                                                                    \end{array}2ip\tau
                                                                                                  \right)\right.
\nonumber\\
&\hspace{6em} +\left.\left.\left(-\frac{p\tau}{2}\right)^{i\mu}\frac{\mathrm{csch}\,\pi\mu}{(2i\mu+1)\Gamma(1+i\mu)}~_2F_2\left(
                                                                                                    \begin{array}{cc}
                                                                                                      1/2+i\mu, & 1/2+i\mu; \\
                                                                                                      3/2+i\mu, & 1+2i\mu; \\
                                                                                                    \end{array}2ip\tau
                                                                                                  \right)\right\}\right|^0_{-\infty}.
\end{align}
The lower limit is zero after we add a small imaginary part to $\tau$ to make the oscillating component exponentially suppressed~\cite{Maldacena:2002vr}, while the upper limit is zero because of the $(-\tau)^{1/2}$ prefactor. Therefore the integrals containing (\ref{IntegralHankelExpExact}) all vanish. Note that this conclusion has nothing to do with our assumption of large mass. The remaining terms, (\ref{Integral03}), (\ref{Integral07}), (\ref{Integral08}), (\ref{Integral09}) and (\ref{Integral10}), will contribute to non-Gaussianity. Let us investigate a typical one
\begin{align}
\left\langle \delta\theta^3 \right\rangle_7 = & -\delta^{(3)}(\bpsum)16R^3\dot\theta_0^3V'''u_{p_1}(0)u_{p_2}(0)u_{p_3}(0)
\nonumber\\
& \times \Re \left[\int^0_{-\infty}d\tau~a(\tau)^4v_{\bp_1}(\tau)v_{\bp_2}(\tau)v_{\bp_3}(\tau) \prod^3_{j=1}\int^\tau_{-\infty}d\tau_j~a(\tau_j)^3v_{\bp_j}^\ast(\tau_j)u'^\ast_{\bp_j}\right]
\nonumber\\
=&(2\pi)^3\delta^{(3)}(\bpsum)\frac{-\pi^3}{32}\frac{\dot\theta_0^3}{HR^3}V'''\frac{e^{-3\pi\mu}}{p_1p_2p_3}
\nonumber\\
&\times \Re \left[ \int^0_{-\infty}d\tau~(-\tau)^{1/2}H_{i\mu}^{(1)}(-p_1\tau)H_{i\mu}^{(1)}(-p_2\tau)H_{i\mu}^{(1)}(-p_3\tau) \right.
\nonumber\\
& \left. \hspace{2em} \times\prod^3_{j=1}\int^\tau_{-\infty}\frac{d\tau_j}{(-\tau_j)^{1/2}}H_{i\mu}^{(1)}(-p_j\tau_j)^\ast e^{ip_j\tau_j} \right] \, .
\label{I7}
\end{align}
We have already used the symmetry to include the permutations by enlarging the integral ranges. The integral (\ref{I7}) can be done analytically only when we use the asymptotic form (\ref{HankelAsympZ<<Mu}) of the Hankel function for $\mu\gg1$ and the integral (\ref{IntegralHankelExp}), both derived in Appendix \ref{app:Hankel}. Using them repeatedly, we can have
\begin{equation}\label{DeltaTheta7}
\left\langle \delta\theta^3 \right\rangle_7=-(2\pi)^3\delta^3(\bpsum)\frac{\dot\theta_0^3}{2\mu^6HR^3}\frac{V'''}{\pprod\left(\psum\right)^3} \, .
\end{equation}
We can proceed more or less the same for the remaining integrals. The results are similar, and adding them together will contribute a factor of 8 to (\ref{DeltaTheta7}). The final result is
\begin{equation}
\left\langle \delta\theta^3 \right\rangle \mapsto -(2\pi)^3\delta^{(3)}(\bpsum) \frac{4\dot\theta_0^3}{\mu^6HR^3} \frac{V'''}{p_1p_2p_3\left( p_1+p_2+p_3 \right)^3} \, .
\end{equation}
Therefore, the three-point function of the curvature perturbation from the heavy vertex is, in terms of the shape function,
\begin{align}
\langle\mathcal{R}(\mathbi{p}_1)\calR(\mathbi{p}_2)\calR(\mathbi{p}_3)\rangle & = (2\pi)^7\delta^{(3)}(\bpsum)\mathcal{P}_\mathcal{R}^2\frac{S_{\delta\sigma^3}(p_1,p_2,p_3)}{p_1^2p_2^2p_3^2} \, ,
\\\label{temp1}
S_{\delta\sigma^3}(p_1,p_2,p_3) & = \frac{4\pi^6R\dot\theta_0^4c_s^2}{\mu^6H^6}V'''\frac{p_1p_2p_3}{\left(p_1+p_2+p_3\right)^3} \, .
\end{align}
Again we preserve a $c_s^2$ factor originate from the definition of the shape function via $\mathcal{P}_\mathcal{R}^2$ as to compare with the result in Section \ref{sec:EFT}. We can see from (\ref{temp1}) that this shape is of equilateral type, and is exactly the same as that the $V'''$-term in the effective action (\ref{ActionEff1}) generates, i.e. the corresponding part of $S_\lambda(p_1,p_2,p_3)$ in (\ref{ShapeEFTbyMass}), except for the difference in the definitions of effective masses. The corresponding non-linear parameter is
\begin{equation}\label{fNLinin}
f_\mathrm{NL}^{\delta\sigma^3}=\frac{40}{243}\frac{R\dot\theta_0^4c_s^2}{H^6\mu^6}V''' \, ,
\end{equation}
which is also the same. The efforts in this section have verified the simple result obtained by the effective theory in the previous section, and therefore we have shown the equivalence of the effective theory and the in-in formalism up to the third order in perturbations.

The different definition of effective masses is negligible when we impose the condition to guarantee the validity of the in-in formalism perturbative series. To make the perturbative series of the exponent in (\ref{def:DeltaTheta^3}) converge, we require
\begin{equation}\label{cond:inin1}
\frac{\dot\theta_0^2}{H^2}\ll1 \, .
\end{equation}
Together with the condition that the mass parameter $\mu$ be large,
\begin{equation}
\mu^2\approx\frac{\M^2}{H^2}=\frac{\m^2}{H^2}+4\frac{\dot\theta_0^2}{H^2}\gg1 \, ,
\end{equation}
we know that $\mu^2\gg1$ implies $\m^2\gg H^2$ when $\dot\theta_0^2\ll H^2$. Therefore, the hierarchy $\dot\theta_0^2\ll H^2\ll\m^2$ gives us
\begin{equation}
c_s^2\approx1 \, ,
\end{equation}
which makes the difference in the definitions of effective mass unimportant.

\section{Discussion}
\label{sec:discuss}

The observational effects of heavy fields during inflation are an interesting issue since inflation is our closest window towards the Planck scale. We believe that if the existence of heavy fields in inflation are verified and identified as the low-energy realization of supergravity or even string landscape, we should have greater insights into the physics of quantum gravity. Many effects have been taken into account as to study the inflationary picture beyond the simplest models, and in this article we have tried to connect some of them. Here we will make in order the main accomplishments and relevant comments.

In a macroscopic point of view, our work is an effort to study the non-Gaussianity originated from the self-interactions of heavy fields during inflation, and some part of the origin of the parameter $M_3$ of the effective action of the Goldstone boson (\ref{ActionGoldstone}). This is done by deriving the effective action (\ref{ActionEff2}) by integrating out the heavy field in the infrared region in a de Sitter background. This can be extended up to higher order, so that if the heavy non-linear interaction terms are purely coming from the potential in the form $\sum_{n\geq3}V^{(n)}\delta\sigma^n/n!$, and in general the coupling $M_n^4$ contains the contribution of $n$-th derivative $V^{(n)}$. We have done some first-step analysis of the expansion to arbitrary order in Appendix \ref{app:n-point}.
It is interesting to note that via the quadratic interaction between the light and heavy fields, higher order derivatives of the heavy field potential are transferred into the expansion parameters of the curvature perturbation, i.e.
\begin{equation}
\calR = \calR_g + \frac{3}{5}\fnl\calR_g^2 + \left( \frac{3}{5} \right)^2\gnl\calR_g^3 + \cdots \, ,
\end{equation}
and we can interpret $V'''\mapsto\fnl$, $V^{(4)}\mapsto\gnl$, and so on. In terms of the Goldstone effective action, this is more apparent. For example, up to quartic order using $\dot{\delta\theta} \approx \dot\theta_0\dot\pi$, we find
\begin{equation}\label{temp2}
S_\eff \supset \int a^3 \mpl^2\dot{H} \left[ \frac{RV'''}{6\m^2} \left( \frac{1}{c_s^2}-1 \right)^2 \dot\pi^3 + \frac{R^2}{16\m^2} \left( \frac{3V'''^2}{\m^2} - V^{(4)} \right) \left( \frac{1}{c_s^2}-1 \right)^3 \dot\pi^4 + \cdots \right] \, ,
\end{equation}
and from the coefficient of $\dot\pi^4$ we can read the non-linear parameters participating in the trispectrum as $V^{(4)}\mapsto\gnl$ and $\left( V^{(3)} \right)^2\mapsto\fnl^2\sim\tnl$.

Besides, we have shown that such non-Gaussianity can be dominant even in the limit $\m\to\infty$, and is possible to be large. An estimation compared to the Planck data~\cite{planck2013} shows that the theoretical constraint is much broader than the observational one, which, as well as the fact that local non-Gaussianity is very small~\cite{Ade:2013ydc}, implies that the assumptions we have set in this article is compatible with observations.

Another goal of this work is to show the equivalence of the effective theory and in-in formalism by investigating an explicit model of the quasi-single field with large mass. This is not completely done, but we have made some firm steps towards this goal.
In this article we have extended a previous work~\cite{Pi:2012gf} up to third order level. It is likely that these two approaches are equivalent up to all orders (without loops) when $\m\to\infty$. As the effective action (\ref{ActionEff2}) does not include the loop corrections of the heavy field, it is also interesting to study the loop effects~\cite{Senatore:2009cf}. Whether the loop corrections will be large or not is still an open question, which is quite interesting, but far beyond the scope of the present work.

\subsection*{Acknowledgments}

We are grateful of Yifu Cai and Jaume Garriga for useful discussions and comments.
SP thanks the GCOE Bilateral International Exchange Program of Kyoto University
and the hospitality of the Yukawa Institute for Theoretical Physics, Kyoto University,
when this work was initiated.
JG and SP acknowledge the Max-Planck-Gesellschaft, the Korea
Ministry of Education, Science and Technology, Gyeongsangbuk-Do and Pohang City
 for the support of the Independent Junior Research Group at the Asia Pacific
Center for Theoretical Physics.
MS is supported in part by JSPS Grant-in-Aid for Scientific Research (A)
No.~21244033.

\appendix

\section{Hankel function with a large complex order}
\label{app:Hankel}

We use the asymptotic form of the Hankel function with a large pure imaginary order $i\mu$~\cite{Watson:1962},
\begin{align}
H_{i\mu}^{(1)}(z) \sim & \sqrt{\frac{2}{\pi}}\frac{z^{i\mu}}{(\mu^2+z^2)^{1/4}} \exp\left\{i\left[\sqrt{\mu^2+z^2}-\mu\log\left(\mu+\sqrt{\mu^2+z^2}\right) -i\frac{\mu\pi}{2}\right]-i\frac{\pi}{4}\right\}
\nonumber\\
&\times\sum_{m=0}^\infty\frac{\Gamma(m+1/2)}{\Gamma(1/2)}\left(\frac{2}{i}\right)^m\frac{A_m}{(\mu^2+z^2)^{m/2}} \, ,
\end{align}
where
\begin{align}
A_0&=1 \,,
\\
A_1&=\frac{1}{8}-\frac{5}{24}\frac{\mu^2}{\mu^2+z^2} \,,
\\
A_2&=\frac{3}{128}-\frac{77}{576}\frac{\mu^2}{\mu^2+z^2}+\frac{385}{3456}\left(\frac{\mu^2}{\mu^2+z^2}\right)^2 \,,
\end{align}
and so on. The form we will use is the asymptotic behavior when $z\ll\mu$,
\begin{align}\label{HankelAsympZ<<Mu}
H_{i\mu}^{(1)}(z) \rightarrow&\sqrt{\frac{2}{\pi\mu}}e^{\pi\mu/2}e^{-i\pi/4}\left(\frac{ez}{2\mu}\right)^{i\mu} \exp \left\{ i\left[ \frac{z^2}{4\mu}-\frac{z^4}{32\mu^3}+\mathcal O\left(\mu^{-5}\right) \right] \right\}
\nonumber\\
&\times\left[1-\frac{1}{12i\mu}-\frac{1}{288\mu^2}-\frac{z^2}{4\mu^2}+\frac{z^2}{4i\mu^3}+\mathcal O(\mu^{-4})\right] \, .
\end{align}
Another useful form is the one with $z\gg\mu$,
\begin{align}\label{HankelAsympZ>>Mu}
H_{i\mu}^{(1)}(z)\rightarrow&\sqrt{\frac{2}{\pi z}}e^{\pi\mu/2}e^{-i\pi/4} \exp \left\{ i\left[ z-\frac{\mu^2}{2z^2}+\mathcal{O}\left(z^{-4}\right) \right] \right\}
\nonumber\\
&\times\left[ 1+\frac{1}{8iz}-\frac{9}{128z^2}-\frac{\mu^2}{4z^2}-\frac{17\mu^2}{24iz^3}+\mathcal{O}(z^{-4})\right] \, .
\end{align}

A direct result from such an asymptotic behavior which is used frequently in the explicit calculations in Section~\ref{sec:in-in} is the integral of the form
\begin{equation}
\int^{\tau'}_{-\infty}\frac{d\tau}{(-\tau)^{1/2}}{H_{i\mu}^{(1)}}^*(-p\tau) e^{ip\tau} \, .
\end{equation}
This can be done by invoking (\ref{HankelAsympZ<<Mu}) to integrate the asymptotic expression,
\begin{align}
& \int^{\tau'}_{-\infty}d\tau~(-\tau)^{-1/2}{H_{i\mu}^{(1)}}^*(-p\tau) e^{ip\tau}
\nonumber\\
& \to \sqrt{\frac{2}{\pi\mu}}e^{\pi\mu/2}\left(\frac{2\mu}{ez}\right)^{i\mu}\left\{\left(1-\frac{1}{12i\mu}-\frac{1}{288\mu^2}\right)\int^{\tau'}_{-\infty} d\tau~(\tau)^{-1/2-i\mu}\exp\left[i\left(p\tau-\frac{p^2\tau^2}{4\mu}\right)\right]\right.
\nonumber\\
&\left. \hspace{10em} -\frac{p^2}{4\mu^2}\int_{-\infty}^{\tau'}d\tau~(-\tau)^{3/2-i\mu}\exp\left[i\left(p\tau-\frac{p^2\tau^2}{4\mu}\right)\right]\right\} \,.
\end{align}
This is valid when $-p\tau\ll\mu$. But since $\mu$ is very large, the contribution from the extreme ultraviolet region is very small. Here and after, we will use an approximate expression of the integral of $x^{a+bi\mu}$ multiplied by an exponent of $x$. This integral, with $\mu\gg a,b,A,B$, is found by first expanding the exponent as an infinite sum as
\begin{equation}
\int d\tau~(-\tau)^{a+ib\mu}e^{i(A\tau+B\tau^2)} \to -\frac{(-\tau)^{a+1+ib\mu}}{ib\mu}e^{i(A\tau+B\tau^2)} \, .
\label{IntegralPolyExp}
\end{equation}
Thus we have
\begin{align}\label{IntegralHankelExp}
\int^{\tau'}_{-\infty}\frac{d\tau}{(-\tau)^{1/2}}{H_{i\mu}^{(1)}}^*(-p\tau) e^{ip\tau} \to &i\sqrt{\frac{2}{\pi\mu^3}}e^{\pi\mu/2}e^{i\pi/4}\left(\frac{2\mu}{ep}\right)^{i\mu}(-\tau')^{1/2-i\mu} \exp\left[i\left(p\tau'-\frac{p^2\tau'^2}{4\mu}\right)\right]
\nonumber\\
&\times \left( 1-\frac{1}{12i\mu}-\frac{1}{288\mu^2}-\frac{p^2\tau'^2}{4\mu^2} \right) \, ,
\end{align}
after we discard the highly oscillating upper limit. Although the phase factor in this expression is complicated, it will be canceled since the integral we are going to do consists of the multiplication of three Hankel functions and their complex conjugates. Similarly, by using (\ref{HankelAsympZ<<Mu}) we can convert every integral of the products of Hankel functions into the form of (\ref{IntegralPolyExp}), and write it in a simple exponential form in the limit $\mu\rightarrow\infty$.

\section{Complete contractions for three point function}
\label{app:Integral}

\allowdisplaybreaks

In Section \ref{sec:in-in} we have encountered a contraction of the expectation value (\ref{def:DeltaTheta^3}). Expanding the exponent up to second order and contracting all the possible combinations, we can find the three-point function of the inflaton fluctuation $\delta\theta$ as a sum of ten terms~\cite{qsfi}
\begin{equation}
\left\langle \delta\theta^3 \right\rangle = 16R^3\dot\theta_0^3V'''(2\pi)^3\delta^{(3)}(\bpsum)\sum_{i=1}^{10} \left\langle \delta\theta^3 \right\rangle_i \, ,
\end{equation}
and each term is given by
\begin{align}
\label{Integral01}
\left\langle \delta\theta^3 \right\rangle_1 = & -
u_{p_1}^* u_{p_2} u_{p_3}(0) \,
\Re \left[ \int_{-\infty}^0 d\tau_1~ a^3   v_{p_1}^*
u'_{p_1}(\tau_1)
\int_{-\infty}^{\tau_1} d\tau~ a^4   v_{p_1}
v_{p_2} v_{p_3}(\tau) \right.
\nonumber\\
& \hspace{7.5em} \times \left.
\int_{-\infty}^0 d\tau_2~ a^3   v_{p_2}^*
u_{p_2}^{\prime *} (\tau_2)
\int_{-\infty}^{\tau_3} d\tau_4~ a^3   v_{p_3}^*
u_{p_3}^{\prime *} (\tau_3) \right] \, ,
\\
\label{Integral02}
\left\langle \delta\theta^3 \right\rangle_2 = & -
u_{p_1}^* u_{p_2} u_{p_3}(0) \,
\Re \left[ \int_{-\infty}^0 d\tau~ a^4   v_{p_1}^*
v_{p_2} v_{p_3}(\tau)
\int_{-\infty}^{\tau} d\tau_1~ a^3   v_{p_1}
u'_{p_1}(\tau_1) \right.
\nonumber\\
& \hspace{7.5em} \times
\left. \int_{-\infty}^0 d\tau_2~ a^3   v_{p_2}^*
u_{p_2}^{\prime *} (\tau_2)
\int_{-\infty}^{\tau_3} d\tau_3~ a^3   v_{p_3}^*
u_{p_3}^{\prime *} (\tau_3) \right] \, ,
\\
\label{Integral03}
\left\langle \delta\theta^3 \right\rangle_3 = &
u_{p_1} u_{p_2} u_{p_3}(0) \,
\Re \left[
\int_{-\infty}^0 d\tau~ a^4   v_{p_1}
v_{p_2} v_{p_3}(\tau)
\right.
\nonumber\\
& \hspace{6.3em} \times \left.
\int_{-\infty}^0 d\tau_1~ a^3   v_{p_1}^*
u_{p_1}^{\prime *} (\tau_1)
\int_{-\infty}^{\tau_1} d\tau_2~ a^3   v_{p_2}^*
u_{p_2}^{\prime *} (\tau_2)
\int_{-\infty}^{\tau_2} d\tau_3~ a^3   v_{p_3}^*
u_{p_3}^{\prime *} (\tau_3)
\right] \, ,
\\
\label{Integral04}
\left\langle \delta\theta^3 \right\rangle_4 = &
u_{p_1}^* u_{p_2} u_{p_3}(0)
\Re \left[
\int_{-\infty}^0 d\tau_1~ a^3   v_{p_1}
u_{p_1}^{\prime} (\tau_1)
\right.
\nonumber\\
& \hspace{6.3em} \times \left.
\int_{-\infty}^0 d\tau~ a^4   v_{p_1}^*
v_{p_2} v_{p_3}(\tau)
\int_{-\infty}^{\tau} d\tau_2~ a^3   v_{p_2}^*
u_{p_2}^{\prime *} (\tau_2)
\int_{-\infty}^{\tau_2} d\tau_3~ a^3   v_{p_3}^*
u_{p_3}^{\prime *} (\tau_3)
\right] \, ,
\\
\label{Integral05}
\left\langle \delta\theta^3 \right\rangle_5 = &
u_{p_1}^* u_{p_2} u_{p_3}(0)
\Re \left[
\int_{-\infty}^0 d\tau_1~ a^3   v_{p_1}
u_{p_1}^{\prime} (\tau_1)
\right.
\nonumber\\
& \hspace{6.3em} \times \left.
\int_{-\infty}^0 d\tau_2~ a^3   v_{p_2}
u_{p_2}^{\prime *} (\tau_2)
\int_{-\infty}^{\tau_2} d\tau~ a^4   v_{p_1}^*
v_{p_2}^* v_{p_3}(\tau)
\int_{-\infty}^{\tau} d\tau_3~ a^3   v_{p_3}^*
u_{p_3}^{\prime *} (\tau_3)
\right] \, ,
\\
\label{Integral06}
\left\langle \delta\theta^3 \right\rangle_6 = &
u_{p_1}^* u_{p_2} u_{p_3}(0)
\Re \left[
\int_{-\infty}^0 d\tau_1~ a^3   v_{p_1}
u_{p_1}^{\prime} (\tau_1)
\right.
\nonumber\\
& \hspace{6.3em} \times \left.
\int_{-\infty}^0 d\tau_2~ a^3   v_{p_2}
u_{p_2}^{\prime *} (\tau_2)
\int_{-\infty}^{\tau_2} d\tau_3~ a^3   v_{p_3}
u_{p_3}^{\prime *} (\tau_3)
\int_{-\infty}^{\tau_3} d\tau~ a^4   v_{p_1}^*
v_{p_2}^* v_{p_3}^*(\tau)
\right] \, ,
\\
\label{Integral07}
\left\langle \delta\theta^3 \right\rangle_7 = & -
u_{p_1} u_{p_2} u_{p_3}(0)
\Re \left[
\int_{-\infty}^0 d\tau~ a^4   v_{p_1}
v_{p_2} v_{p_3}(\tau)
\int_{-\infty}^{\tau} d\tau_1~ a^3   v_{p_1}^*
u_{p_1}^{\prime *} (\tau_1)
\right.
\nonumber\\
& \hspace{7.5em} \times \left.
\int_{-\infty}^{\tau_1} d\tau_2~ a^3   v_{p_2}^*
u_{p_2}^{\prime *} (\tau_2)
\int_{-\infty}^{\tau_2} d\tau_3~ a^3   v_{p_3}^*
u_{p_3}^{\prime *} (\tau_3)
\right] \, ,
\\
\label{Integral08}
\left\langle \delta\theta^3 \right\rangle_8 = & -
u_{p_1} u_{p_2} u_{p_3}(0)
\Re \left[
\int_{-\infty}^0 d\tau_1~ a^3   v_{p_1}
u_{p_1}^{\prime *} (\tau_1)
\int_{-\infty}^{\tau_1} d\tau~ a^4   v_{p_1}^*
v_{p_2} v_{p_3}(\tau)
\right.
\nonumber\\
& \hspace{7.5em} \times \left.
\int_{-\infty}^{\tau} d\tau_2~ a^3   v_{p_2}^*
u_{p_2}^{\prime *} (\tau_2)
\int_{-\infty}^{\tau_2} d\tau_3~ a^3   v_{p_3}^*
u_{p_3}^{\prime *} (\tau_3)
\right] \, ,
\\
\label{Integral09}
\left\langle \delta\theta^3 \right\rangle_9 = & -
u_{p_1} u_{p_2} u_{p_3}(0)
\Re \left[
\int_{-\infty}^0 d\tau_1~ a^3   v_{p_1}
u_{p_1}^{\prime *} (\tau_1)
\int_{-\infty}^{\tau_1} d\tau_2~ a^3   v_{p_2}
u_{p_2}^{\prime *} (\tau_2)
\right.
\nonumber\\
& \hspace{7.5em} \times \left.
\int_{-\infty}^{\tau_2} d\tau~ a^4   v_{p_1}^*
v_{p_2}^* v_{p_3}(\tau)
\int_{-\infty}^{\tau} d\tau_3~ a^3   v_{p_3}^*
u_{p_3}^{\prime *} (\tau_3)
\right] \, ,
\\
\label{Integral10}
\left\langle \delta\theta^3 \right\rangle_{10} = & -
u_{p_1} u_{p_2} u_{p_3}(0)
\Re \left[
\int_{-\infty}^0 d\tau_1~ a^3   v_{p_1}
u_{p_1}^{\prime *} (\tau_1)
\int_{-\infty}^{\tau_1} d\tau_2~ a^3   v_{p_2}
u_{p_2}^{\prime *} (\tau_2)
\right.
\nonumber\\
& \hspace{7.5em} \times \left.
\int_{-\infty}^{\tau_2} d\tau_3~ a^3   v_{p_3}
u_{p_3}^{\prime *} (\tau_3)
\int_{-\infty}^{\tau_3} d\tau~ a^4   v_{p_1}^*
v_{p_2}^* v_{p_3}^*(\tau)
\right] \, ,
\end{align}
with a summation over five different permutations of $p_1$, $p_2$ and $p_3$ for each term.

\section{Structure of effective action with non-linear interactions}
\label{app:n-point}

Here we consider a heavy field with arbitrarily non-linear self-interactions. To obtain the effective action of the curvature perturbation, we have to eliminate the heavy field $\delta\sigma$ by solving
\begin{equation}\label{eom:DeltaSigmaAccurate}
\frac{\partial V(\sigma_0+\delta\sigma)}{\partial(\delta\sigma)}-(\dot\theta_0^2-2\dot\theta_0\dot{\delta\theta}-\dot{\delta\theta}^2)\delta\sigma
=2R\dot\theta_0\dot{\delta\theta}+R\dot{\delta\theta}^2 \, .
\end{equation}
This is the non-linear extension of (\ref{eom:DeltaSigma}). To solve this equation perturbatively, we first take the Taylor expansion of $V(\sigma_0+\delta\sigma)$ around $\delta\sigma=0$. Noting that its second order derivative $V''\equiv\m^2+\dot\theta^2$ is the dominant term on the left hand side of (\ref{eom:DeltaSigmaAccurate}), we can solve the equation up to an arbitrary order by using the formula derived in \cite{Lewis:1935}:
\begin{align}\label{sol:DeltaSigmaAccurate}
\delta\sigma = & \sum_{p_2=0}^\infty \sum_{p_3=0}^\infty \cdots \sum_{p_n=0}^\infty (-1)^{p_2+p_3+\cdots+p_n} \left(\frac{V^{(3)}}{2!}\right)^{p_2} \left(\frac{V^{(4)}}{3!}\right)^{p_3} \cdots \left(\frac{V^{(n+1)}}{n!}\right)^{p_n}
\nonumber\\
& \times \frac{(2p_2+3p_3+\cdots+np_n)!}{(1+p_2+2p_3+\cdots+(n-1)p_n)!}
\frac{\left(2R\dot\theta_0\dot{\delta\theta}+R\dot{\delta\theta}^2\right)^{1+p_2+2p_3+\cdots+(n-1)p_n}}
{\left(\m^2-2\dot\theta_0\dot{\delta\theta}-\dot{\delta\theta}^2\right)^{1+2p_2+3p_3+\cdots+np_n}} \, ,
\end{align}
where $n$ is the highest order which the slope of the potential $V'(\sigma_0+\delta\sigma)$ is expanded to, and $V^{(i)}$ denotes the $i$-th derivative of the potential $V(\sigma)$ with respect to $\sigma$ evaluated at $\sigma_0$. For a specific $n$, (\ref{sol:DeltaSigmaAccurate}) is an infinite series of $\dot{\delta\theta}$, and the terms can be categorized by perturbative order $(1+p_2+2p_3+\cdots+(n-1)p_n)$. That is, to reorganize the infinite series (\ref{sol:DeltaSigmaAccurate}) by the order of $\dot{\delta\theta}$. We define the term proportional to $\dot{\delta\theta}^j$ as $\delta\sigma^{(j)}$ for an arbitrary integer $j\leq n$, and a perturbative solution to $\delta\sigma$ is a sum of $\delta\sigma^{(j)}$. The leading order term $\delta\sigma^{(1)}$, of course, is the case when all the $p_j$'s are zero, which gives
\begin{equation}\label{DeltaSigmaLinear}
  \delta\sigma^{(1)}=\frac{2R\dot\theta_0\dot{\delta\theta}+R\dot{\delta\theta}^2}{\m^2-2\dot\theta_0\dot{\delta\theta}-\dot{\delta\theta}^2}
  \approx\frac{2R\dot\theta_0}{\m^2}\dot{\delta\theta}.
\end{equation}
The next-to-leading order is $p_2=1$ and $p_i=0$ when $i>2$\footnote{As we have demonstrated in the Section~\ref{sec:EFT}, the terms not involving $V'''$ are suppressed by heavy mass and negligible.},
\begin{equation}\label{DeltaSigmaBilinear}
  \delta\sigma^{(2)}\approx-\frac{V^{(3)}}{2}\frac{\left(\delta\sigma^{(1)}\right)^2}{\m^2} \, ,
\end{equation}
which together with (\ref{DeltaSigmaLinear}) replicate our constraint equation (\ref{Constraint}) used in Section~\ref{sec:EFT}.

For a generic $n$, the solution in (\ref{sol:DeltaSigmaAccurate}) is accurate up to $n$-th order, and the solution reorganized by the perturbative order $n$ can be determined by solving the linear Diophantine equation
\begin{equation}\label{Diophantine}
  p_2+2p_3+\cdots+(n-1)p_{n}=n-1 \, .
\end{equation}
Every solution set of (\ref{Diophantine}) contributes to one independent coefficient to $\delta\sigma^{(n)}$, and according to the convergence of the series of potential $V(\sigma_0+\delta\sigma)$, (\ref{Vseries}), all of the terms are of equal importance. For example, for $p_n=1$ and other $p_i=0$,
\begin{equation}\label{DeltaSigmaN1}
  \delta\sigma^{(n)}\supseteq-\frac{V^{(n+1)}}{n!}\frac{\left(\delta\sigma^{(1)}\right)^n}{\m^2} \, .
\end{equation}
Another typical term corresponding to solution $p_{n-1}=1$, $p_2=1$ and other $p_i=0$ is
\begin{equation}\label{DeltaSigmaN2}
  \delta\sigma^{(n)}\supseteq\frac{V^{(3)}}{2!}\frac{V^{(n)}}{(n-1)!}(n+1) \frac{\left(\delta\sigma^{(1)}\right)^{n+1}}{\m^4} \, .
\end{equation}
It has almost the same upper limit as (\ref{DeltaSigmaN1}) if we use (\ref{est:pertvalid}) to estimate $V^{(3)}$. So we see that there are no hierarchies among all the terms in the coefficient of $\delta\sigma^{(n)}$, and the their contributions may be equally important in principle.

To obtain an accurate effective action up to $n$-th order, one can substitute (\ref{sol:DeltaSigmaAccurate}) back into the Lagrangian (\ref{ActionPert}), which is too complex to be displayed here. But we can also see that the coefficient of $\dot{\delta\theta}^n$ term consists of $V^{(n)}$ and the products of lower derivatives, which are in principle of equal magnificence. In Section \ref{sec:discuss} we displayed the result with $n=4$ as an example.

\end{document}